# Acoustic measurements above a plate carrying Lamb waves

Andreas Sørbrøden Talberg [1], Tonni Franke Johansen [1,2]

[1] Norwegian University of Science and Technology, NTNU
[2] SINTEF ICT
Contact email: andreasstalberg@gmail.com

## Abstract

This article presents a set of acoustic measurements conducted on the Statoil funded Behind Casing Logging Set-Up, designed by SINTEF Petroleum Research to resemble an oil well casing. A set of simple simulations using COMSOL Multiphysics were also conducted and the results compared with the measurements. The experiments consists of measuring the pressure wave radiated of a set of Lamb waves propagating in a 3 mm thick steel plate, using the so called pitch-catch method. The Lamb waves were excited by a broadband piezoelectric immersion transducer with center frequency of 1 MHz. Through measurements and analysis the group velocity of the fastest mode in the plate was found to be 3138.5 m/s. Measuring the wave radiated into the water in a grid consisting of 8x33 measuring points, the spreading of the plate wave normal to the direction of propagation was investigated. Comparing the point where the amplitude had decreased 50 % relative to the amplitude measured at the axis pointing straight forward from the transducer shows that the wave spread out 3.2 mm after propagating 140 mm in the plate.

## 1 Introduction

According to the Norwegian Petroleum Directorate there have been drilled more than 1500 exploration wells and more than 4200 development wells on the Norwegian Continental Shelf since the discovery of oil in 1966 [1]. Due to the severity of uncontrolled hydrocarbon migration from oil or gas bearing reservoirs to the surroundings, both during production as well as when the well has been plugged for abandonment, it is crucial to cement the well properly. After a cement job has been completed, a proper inspection has to be conducted to evaluate the cement sheath. The operation of placing a permanent plug in a well that has reached the end of its lifetime on offshore exploration wells in Norway, can contribute with 25% of the total drilling costs [2]. Due to the extra cost for the responsible company and the environmental consequences of an improperly executed cement job, the field of cement job evaluation is of great interest.

### Well Cementing

When placing a cement sheath, different criteria have to be fulfilled. The main objective of primary cementing, the stage where cement is placed in the annulus between the casing and the borehole wall, is to create zonal isolation. This is achieved by creating a hydraulic seal between both the cement and borehole wall, as well as between the casing and the cement. There should also be no fluid channels inside the cement [3]. Figure 1 shows different problems that may occur when a cement sheath is placed improperly in a wellbore. Causes of a poorly conducted cement job can be divided into two categories: flow problems of mechanical origin like poor centralization and washouts, and degradation of the cement during the curing stage where the cement may be polluted by fluid or gas [4]. The production capacity of a well may never reach its full potential without complete isolation in the wellbore. To control if the cement sheath has been placed properly, a cement job evaluation has to be conducted.

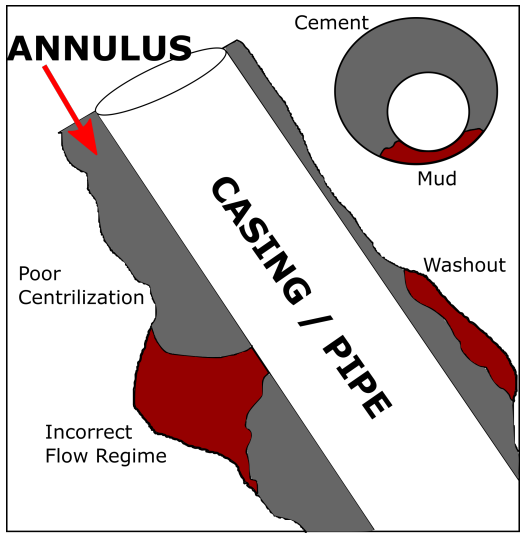

Figure 1: Examples of cement slurry displacement problems in the annulus.

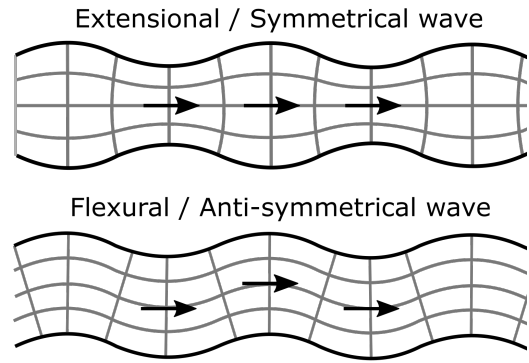

Figure 2: Illustration of the particle displacement in Lamb wave modes. The arrows indicate the direction of the wave propagation.

## Cement Job Evaluation: The Pitch-Catch Technique

Since the 1960s when Zemanek and Caldwell first applied the ultrasonic pulse-echo technique to well logging with their tool named the Borehole Televiewer (1969), the ultrasonic tools have been an important tool in the field of cement job evaluation. The ultrasonic tools usually operate in the frequency range of 200 to 700 kHz, much higher than the acoustic tools. The frequencies are chosen to coincide with the ressonance frequency of the casing [5]. The so called pitch-catch (P-C) method used e.g. by Schlumberger in their Isolation Scanner (2005) [6, 7] is the technique used in this paper.

The concept of the pitch-catch method involves the Lamb wave and the conditions for the following derivations are that the plate, in which the Lamb waves are excited, is an infinitely unbounded plate lying in the $xy$-plane. For sufficiently high frequencies, 80 kHz or more, the pulse from the ultrasonic transducer interacts with a small area of the casing, making the approximation of treating the part of the casing as a part of an infinitely unbounded plate appropriate [8]. A plate supports two infinite sets of Lamb wave modes whose velocities in the plate depend on the relationship between plate thickness, $d$, and the frequency, $f$. The number of possible modes and the velocity of the different modes are decided by the quantity frequency-thickness, $f \cdot d$. As $f \cdot d \to 0$ only the lowest anti-symmetrical mode, $a_0$, and the lowest order symmetrical mode, $s_0$, can appear in the plate. These lowest order modes are called the flexural and the extensional mode respectively and are shown in figure 2. As $f \cdot d$ increases higher order symmetric and anti-symmetric Lamb wave modes may be excited in the plate.

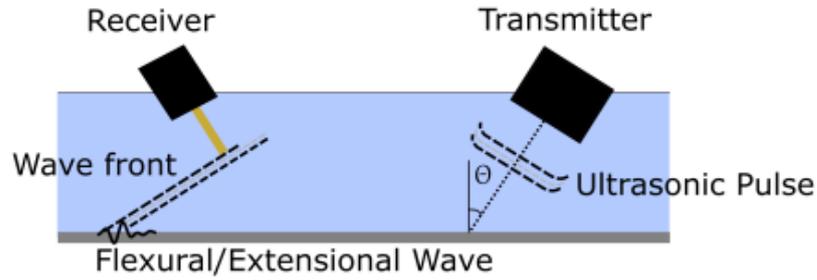

Figure 3: Illustration of the pitch-catch principle. A transmitter emits an ultrasonic pulse which excites Lamb waves in the plate. The Lamb waves propagates in the plate, radiating/leaking energy into the fluid which is detected by the receiver.

In the P-C method, a piezoelectric immersion transducer emits an ultrasonic pulse into the casing fluid. As the pulse hits the solid plate, various Lamb wave modes are excited, dependent on plate thickness, the pulse frequency and the incident angle of the pulse. The Lamb waves then propagate along the plate, radiating or leaking (hence often called Leaky Lamb Waves, LLW) energy into the casing fluid. Due to the velocity of the plate wave being faster than the sound velocity in the casing fluid, the propagating Lamb wave works as an ultrasonic sound source, generating a wave front in the fluid. The wave front in the fluid is then detected by a receiver. A simple sketch of the P-C method is shown in figure 3.

## 2 Set-Up and Experimental Method

**Experimental Set-Up**

The ultrasonic experiments were conducted at the Laboratory of Formation Physics at SINTEF Petroleum Research in Trondheim. SINTEF's Statoil funded "Behind Casing Logging Set-Up" (BeCaLoS, fig. 4) was made available to the author to conduct the ultrasonic experiments on. The BeCaLoS was designed to replicate an oil well casing with the possibility to change casing fluid, the thickness and number of plates, and the annulus material. It is formed as a cylinder with a diameter of 500 mm and a length of 200 mm. In this specific experiment a 3 mm thick steel plate was placed on top of the cylinder. To hold the steel plate to the cylinder and to work as a fluid tank, an open cylinder also with a diameter of 500 mm, was bolted to the first cylinder with a rubber ring placed between the plate and the upper cylinder to keep the tank from leaking. On top of the BeCaLoS a transmitter/receiver holder array was placed to make the positioning of the transmitter and the receiver relatively simple. Positions in $x$-, $y$- and $z$-direction as well as the transmitter/receiver tilt in the $xz$-plane were adjustable parameters in the experiment. Figure 4 shows a picture of the BeCaLoS together with two sketches. In the experiment an ultrasonic pulser-receiver of the type Olympus Panametrics-NDT 5900PR was used to send an electric pulse to a piezoelectric transducer. The transducer used was a Panametrics C302 Immersion Transducer which has a circular aperture with diameter 1 in. (25.4 mm) and emits an ultrasonic broadband pulse with center frequency 1 MHz. The transducer was mounted in the transducer/hydrophone holder array of the BeCaLoS together with a HNR-0500 needle hydrophone from Onda Corp. The hydrophone has a circular aperture with diameter 500 $\mu$m. The electric signal from the hydrophone was sent via the pulser-receiver which amplified the signal, to a digital sampling oscilloscope of the type DSO-X 4024A Oscilloscope from Agilent Technologies. The annulus material was air and the casing fluid was water during the following set of measurement.

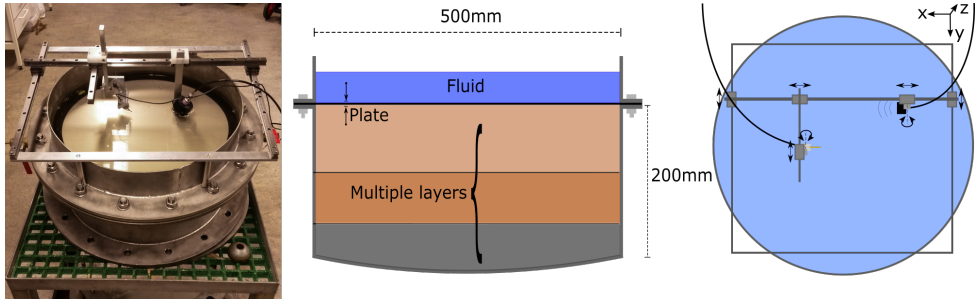

Figure 4: *Left*: SINTEF's Statoil funded Behind Casing Logging Set-Up. *Middle*: Sketch showing the different possibilities regarding number of plates, casing fluid, and annulus material. *Right*: Sketch showing the BeCaLoS viewed from above together with the transducer/hydrophone holder array.

### Experimental Method

A considerable part of the work went to configure the set-up, conducting a set of preliminary measurements to find the optimal tilt of both the transducer and the hydrophone, and to eliminate sources of error. The experimental work conducted on the BeCaLoS was inspired by the work done by K. Hoel [9]. The main measurements which will be presented in the following section consisted of adjusting the *x*- and *y*-position (defined in fig. 4) of the hydrophone with the transducer held still. The *x*-positions ranged from 60 - 200 mm with 20 mm intervals between each measurement point, while the *y*-positions ranged from 0 to 80 mm with 2.5 mm intervals. For each of these measurement points, the pulse detected by the hydrophone was logged using a LabView program. The rest of the configurations during the measurements can be found in Appendix A.

### COMSOL Simulations

COMSOL Multiphysics was used to make a model of the conducted experiment. As shown in figure 5 and 6 the wave from the transducer was modeled as a circular pulse with the same diameter as the transducer (25.4 mm) and with an incident angle of 30°, bounded spatially by a raised cosine function. The incoming was applied as a boundary load to the plate with two periods of a 1 MHz pulse, bounded by a Gaussian window. The modulated plate was a 30 mm wide, 150 mm long and 3 mm thick steel plate in vacuum. The amplitude of the plate displacement as a function of time and space was analyzed, giving information about the waves propagating in the plate.

## 3  Results

The main measurements resulted in a data set consisting of 264 (8x33) pressure pulses detected as a function of time. Eight examples of the data gathered are shown in figure 7 for the measurements along the *x*-axis, together with points indicating the maximum value of the envelope for each measurement (the *x*-axis is defined as the propagation direction of the LLW, illustrated in figure 4). In figure 8 the group velocity of the wave in the plate, found by measuring the first pulse detected by the hydrophone at each measuring point, is plotted together with a mean value. The average group velocity of the first wave was measured to 3138.5 m/s. Comparing this experiment with the work done by Zeroug and Froelich [10] on a steel plate in water, a group velocity of 3100-3200 m/s was expected for the flexural mode.

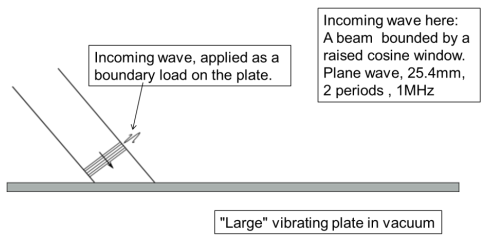

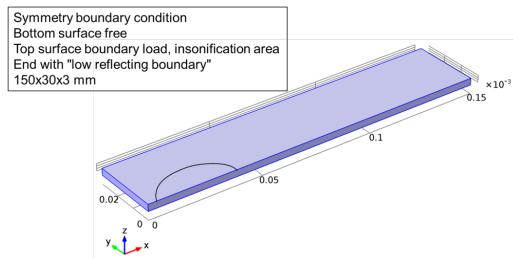

Figure 5: A sketch of the incoming wave modeled as a boundary load on the upper surface of the 3D model.

Figure 6: Image from COMSOL Multiphysics showing a 3D plate in vacuum. The ellipse on the plate indicates where the boundary load representing the incoming wave were applied.

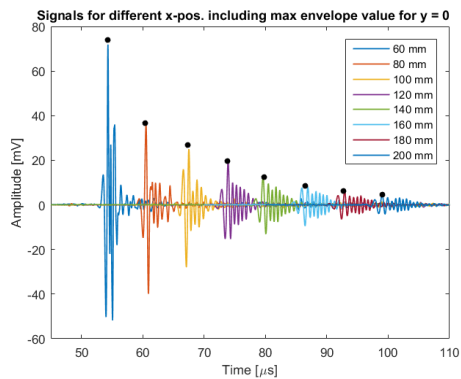

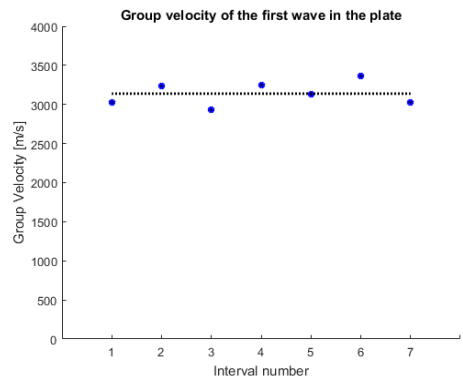

Figure 7: Eight pulses measured by the hydrophone at different $x$-positions. The black dots indicates the maximum of the envelope.

Figure 8: Group velocity of the first Lamb wave as a function of interval number. The black line indicates the average of 3138.5 m/s.

The peak amplitude for each measuring point are shown in a logarithmic, normalized plot in figure 9. As the figure indicates, the attenuation of the LLWs are high due to the radiation of energy into the fluid that is in contact with the plate. In figure 10 the absolute values of the peak amplitudes for $y = 0$ is plotted together with an exponential fit. From the fit an attenuation factor of 0.1927 dB/mm was found. Investigating further how the beam spreads out in the $y$-direction, the peak amplitude for each measuring point was normalized against the peak value at the given $x$-position and $y = 0$. The normalized peak values are shown in a logarithmic plot in figure 11, giving information about the beam profile in the plate. Four cross sections of the beam profile plot with amplitude as a function of $y$ are shown in figure 12. The peak voltage values at $y \gtrsim 50$ mm are mainly noise, which becomes more significant at larger $x$-values due to the normalization against the exponentially decreasing peak amplitude value at the axis. As both the cross section plot and the black lines in the beam profile plot shows, the LLWs does not spread out much in the $y$-direction. After propagating 140 mm in the $x$-direction the -6 dB line, where the amplitude have decreased approximately 50% relative to the corresponding peak value at

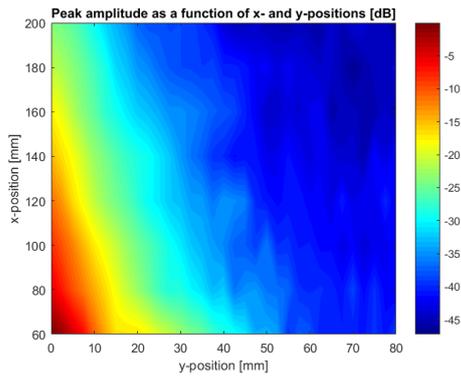

Figure 9: Logarithmic plot of the absolute value of the peak amplitude as a function of the hydrophone's *x*- and *y*-position.

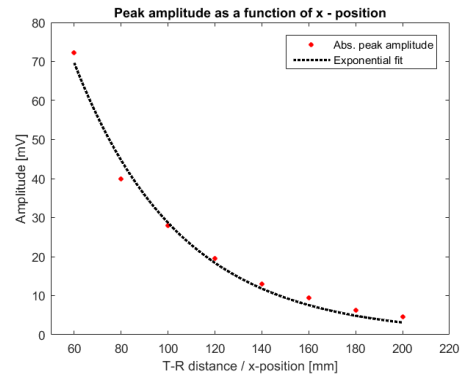

Figure 10: Peak amplitude of the pulse measured along the *x*-axis with an exponential fit with an decaying factor of -0.1927 dB/mm.

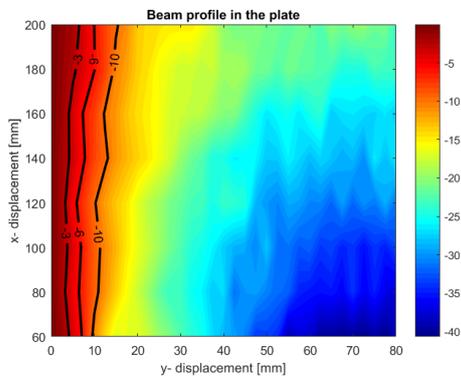

Figure 11: Logarithmic plot of the peak amplitudes normalized against the peak values at *y* = 0. The -3, -6 and -10 dB lines are drawn in the figure.

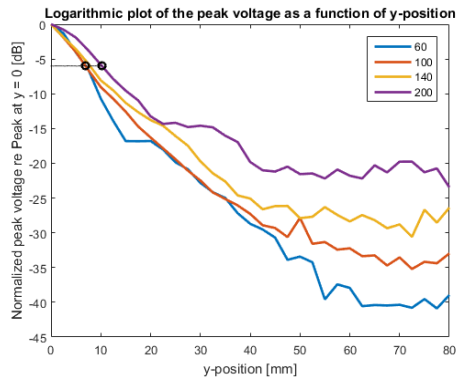

Figure 12: Cross sections of the beam profile figure for four *x*-values. The black dotted line and the two circles indicates the -6 dB points.

*y* = 0 has only moved 3.2 mm in the *y*-direction. This is also indicated by the two black circles in figure 12.

As the Lamb waves propagate in the plate, the different modes propagate with different velocities. In figure 7 eight pulses detected by the hydrophone are plotted, were the pulse becomes wider as it propagate in the plate. In figure 13 the normalized signals for all the 33 *y*-values at *x* = 100 mm are plotted. The red and the blue dotted lines shows two detected pulses radiated of the LLWs traveling at different velocities, indicating that it is two different modes. In figure 14 the time delay of the first detected pulse relative to the pulse detected at *y* = 0 is plotted as a function of *y* for the eight different *x*-positions. The same is shown in figure 15 for the second detected pulse. Figure 16 shows the time delay of the first detected pulse, zoomed in on $y \leq 25$ mm, which shows that the time delay at $y \leq 10$ mm is equal at all of the eight *x*-positions. The radius of the transducer was 12.7 mm and hence a flat curve indicating no time delay could be expected for *y*-values smaller than 12.7 mm. The small time delay for $y \leq 10$ mm may originate from the transducer not emitting a plane ultrasonic pulse, but a spatially bounded, slightly curved pulse. In figure 17 and 18 the time of arrival of the first and second detected pulse respectively are shown as a function of the *x*- and *y*-position of the hydrophone. The figures shows clearly how the second pulse travels slower than the first pulse, indicated by the black lines drawn

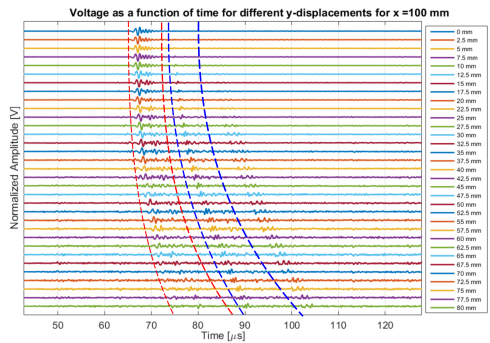

Figure 13: The 33 normalized pulses detected at $x$ = 100 mm. The red and blue dotted lines highlights the first and second detected pulse.

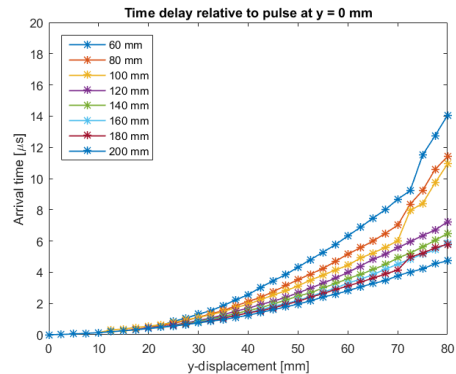

Figure 14: Time delay of the first detected pulse relative to the pulse measured at $y$ = 0 as a function of $y$.

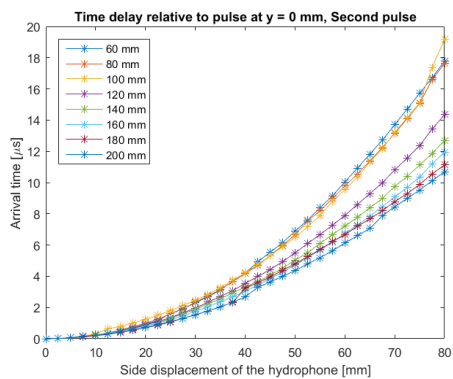

Figure 15: Time delay of the second detected pulse relative to the pulse measured at $y$ = 0 as a function of $y$.

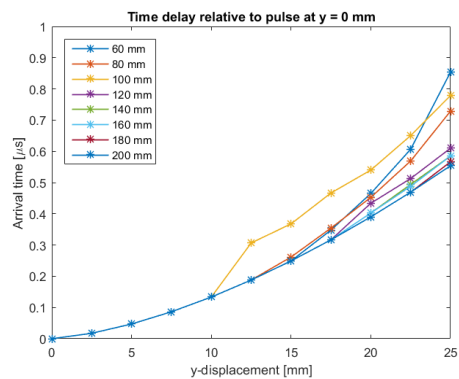

Figure 16: Time delay of the first detected pulse relative to the pulse measured at $y$ = 0 as a function of $y$, zoomed in on smaller $y$-values from fig. 14.

for every 20 $\mu$s.

## COMSOL Simulations

Some results of the COMSOL Multiphysics model of a 3D plate in vacuum are shown in figure 19 and 20. In figure 19 it is clearly shown that the Lamb waves have split into two wave packages traveling at different velocities. The first is an anti-symmetrical mode, while the second is a symmetrical mode. This coincides with the measurements done, where the Lamb waves split into multiple wave packages which travels at different velocities. The fact that the first wave package traveling in the plate seems to be the flexural mode in the simulations and that the expected velocities of the flexural wave in a plate contact with water from Zeroug and Froelich's experiment were 3100-3200 m/s [10], fits with the experiment conducted. Figure 20 shows a snapshot in time of the normalized displacement of the top surface for ten different $y$-positions. The maximum of the second wave package has traveled a shorter $x$-distance for larger $y$-values, relative to the corresponding maximum at $y$ = 0 mm, than the maximum of the first wave package. This coincides well with the plots in figure 14 and 15, indicating the time delay between the maximums of the first and second pulse detected, showing a steeper curve for the second detected pulse.

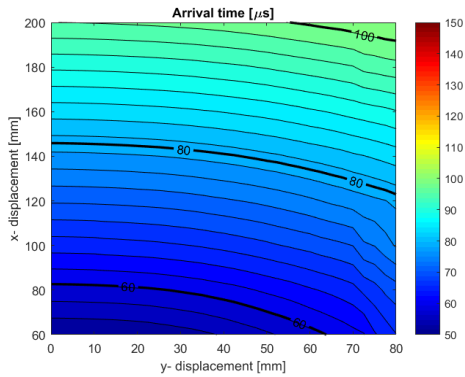

Figure 17: Time of arrival for the first pulse detected by the hydrophone as a function of $x$- and $y$-positions.

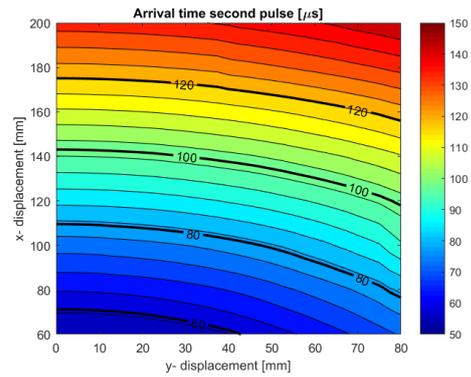

Figure 18: Time of arrival for the second pulse detected by the hydrophone as a function of $x$- and $y$-positions.

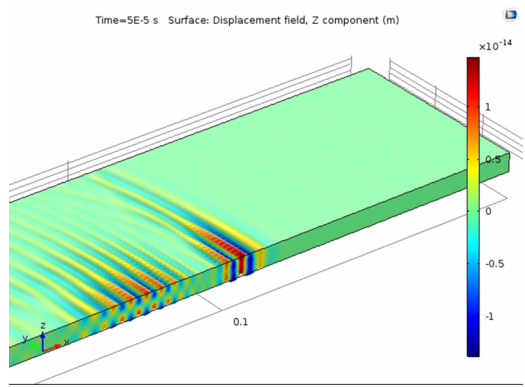

Figure 19: Snapshot of the plate displacement at $t = 50$ $\mu$s. First wave package is anti-symmetrical while the second wave package is symmetrical.

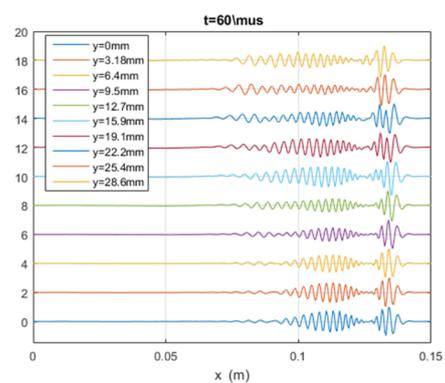

Figure 20: Snapshot of the normalized displacement of the upper surface of the plate for ten different $y$-positions, captured at $t = 60$ $\mu$s.
NB! Legend and plots are in opposite order - the lowest plot belongs to the upper legend entry.

## 4 Conclusion

After a preliminary set of adjustments and measurements the experiments conducted on the BeCaLoS gave good information about the Lamb waves in the plate. The measured group velocity of the Lamb wave are in agreement with the simulations and similar experiments conducted. Inspecting how the plate wave widens in $y$-direction due to the impact of the vertically polarized shear waves, it is shown that the Lamb wave spreading is very small and that the plate wave behaves almost as a soundbeam.

## Nomenclature

**BeCaLoS**    Behind Casing Logging Set-Up
**LLW**    Leaky Lamb Wave
**P-C**    Pitch-Catch

# A  Appendix

Table 1: The configuration of the BeCaLoS during the main measurements.

| Variable | Value |
| --- | --- |
| Transmitter - Receiver distance ($x$-pos.) | 60 - 200 mm |
| Side-/$y$-displacement | 0 - 80 mm |
| Transducer Tilt | 30° |
| Transducer Height | 30 mm |
| Hydrophone Tilt | 33° |
| Hydrophone Height | 30 mm |
| BeCaLoS Tilt | 8.3° |
| Pulse/Receiver Settings | |
| Energy | 32 $\mu$J |
| Gain | 40.0 dB |
| Attenuators | 0.0 dB |